\documentstyle[aps,twocolumn]{revtex}
\begin{document}
\preprint{USACH-FM-00-08 }
\title{QUANTUM GRAVITY AT VERY HIGH ENERGIES}
\author{J. GAMBOA\thanks{E-mail: jgamboa@lauca.usach.cl}and  F.
MENDEZ\thanks{E-mail:
fmendez@lauca.usach.cl}}
\address{Departamento de F\'{\i}sica, Universidad de Santiago de Chile,
Casilla 307, Santiago 2, Chile}
\maketitle
\begin{abstract}
The problem of time and the quantization of three dimensional gravity in the strong coupling regime is studied following path
integral methods. The time is identified with the volume of spacetime. We show that the effective action describes an infinite set of
massless relativistic particles moving in a curved three-dimensional target space, {\it i.e.} a tensionless $3$-brane on
a curved background. If  the cosmological constant  is zero the target space is flat and there is no \lq \lq graviton" propagation ({\it
i.e.}, $G[g_{ij} (2), g_{ij} (1)] = 0$). If the cosmological constant is different from zero, $3D$ gravity is both  classical and
quantum mechanically soluble. Indeed, we find the following results: i) The general exact solutions of the Einstein equations
are singular at $t=0$ showing the existence of a big-bang in this regime and ii) the propagation amplitude between two geometries
$<g_{ij} (2), t_2\vert g_{ij} (1), t_1>$ vanishes  as $t \rightarrow 0$, suggesting that big-bang is suppressed quantum mechanically.
This result is also valid for  $D>3$.
\end{abstract}
\narrowtext

\section{Introduction}

The quantization of the gravitational field is a problem that has
resisted in spite of intense research in the last forty
years \cite{isham}. The conceptual problems involved in the dynamics of
gravity are main barrier towards understanding the gravitational
field at the quantum level. When gravity is present, the role played by the
observer is intrinsic and the concept of time is lost. This last fact
implies that the evolution of the physical states, as it is generally assumed
in quantum mechanics, becomes meaningless and one should try to understand
the dynamics of the gravitational field in some other way. In the past,
several attempts have been tried to adapt the interpretation of quantum
mechanics in the presence of gravitation \cite{thoof}.
These attempts include incorporating the observer, or dropping the
requirement of unitarity in the quantum theory\cite{hawking1}.

The solutions to these difficulties present new and formidable problems in
the quantum treatment of the fulfledged theory. This underlines the
importance of studying simplified models in order to understand the key
features of more complicated four dimensional problems. Along this research
line, in recent years models of gravity in two and three dimensions have
attracted considerable interest. Although these models do not attempt to
describe the real world, they  keep many of the features occurring in four
dimensions, which could really be studied in the quantum version
of simplified models.

Gravity in two dimensions with cosmological constant has been exactly quantized as a
conformal field
theory\cite{poly} and two-dimensional models describing black holes or
cosmological singularities have been found \cite{various}. Although
these models still present difficulties, there is a good indication that
the singularities predicted by the classical theory should be smoothed out
by quantum effects.

In three dimensions similar properties have also been described\cite{deser}. For
instance, a black hole solution\cite{btz} similar to the Kerr one has been
found \cite{carlitos}. Also, an equivalence between $3D$ gravity and
Chern-Simons theory has been noted \cite{pk,witten} but, to our knowledge, no
quantitive argument describing smoothing of singularities at the quantum
level has been given so far.

This kind of argument could be useful in the understanding of the role played  by the
classical singularities in the
quantum theory. There are two prominent examples in the history of physics about
it. One of them is the stability of the hydrogen atom: the
electron never falls to the nucleus because the hamiltonian is self-adjoint and the
current of probability  vanishes in $r=0$. The second example is the smoothing of the electron self-energy by
renomalization in quantum electrodynamics.

In order to understand conceptual issues such as time, probability and evolution of the
physical states in \cite{jg} we proposed an approach to  $2D$
quantum dilaton gravity where the temporal component of the $\phi^A$ two vector -
that defines the target space - play de role of time. Then, if one assumes this point of view one can
prove that indeed there is no big-bang singularity.

The purpose of the present paper is to report new results for quantum gravity in
the strong coupling regime.  Our main results can be summarized as follow; i) The
Einstein equations in the strong coupling regime in 2+1 dimensions are
solved exactly for the topology $\Re \times \Sigma$, where $\Sigma$ is a
Riemann surface and it is shown that there is a cosmological
singularity at $t=0$. ii) The quantum dynamics of this system is
equivalent to an infinite set of relativistic massless particles moving in
three dimensional spacetime (tensionless  3-brane). iii) For any D, we show that for ($\Re \times \Sigma^{D-1}
$), the
initial big-bang singularity is generic and smoothed out by
quantum mechanics.

The paper is organized as follows. In section 2 the physical
meaning of the strong coupling limit is discussed and the necessary
arguments for the gauge fixing are given. In section 3, the Einstein's
equations in the proper-time gauge are solved. In section 4, $3D$ gravity
is quantized in the strong coupling limit using path integral methods and
section 5 contains the conclusions.

\section{The Strong Coupling Limit and the Gauge Fixing Procedure}

The action of gravity for a $D$-dimensional spacetime
is
\begin{equation}
S = \int d^Dx ( \pi^{ij} {\dot g_{ij}} - N {\cal H} - N^i {\cal H}_i ),
\label{action}
\end{equation}
where the constraints are defined as
\begin{eqnarray}
{\cal H} &=& G_{ijkl} \pi^{ij} \pi^{kl} + \frac{1}{G^2} g (
R^{(D-1)} + 2\Lambda),\nonumber \\
{\cal H}_i &=& -2 \pi^j_{i;j},
\label{constr}
\end{eqnarray}
being $G$ the Newton constant,
$\Lambda$ the cosmological constant and $G_{ijkl}$ the
supermetric defined as

\begin{equation}
G_{ijkl} = \frac{1}{2} (g_{ik}g_{jl} + g_{il}g_{jk} -
\frac{2}{D-2} g_{ij}g_{kl}).
\label{smetric}
\end{equation}

The strong coupling limit is equivalent to taking the limit
$G \rightarrow \infty$ and corresponds to replace
${\cal H}$ in (\ref{constr}) by\cite{ish}
\begin{equation}
{\cal H} = G_{ijkl} \pi^{ij} \pi^{kl} + \lambda g,
\label{h1}
\end{equation}
where $\lambda = 2\Lambda/G^2$  being held fixed.  In this limit the spatial
derivatives of the metric are irrelevant and therefore the situation is
equivalent to assuming that the metric of spatial sections only depends
on time.

In this regime the system (\ref{h1}) describes an
infinite set of massless relativistic particles moving in a three
dimensional curved spacetime with metric $G_{AB}$. The additional term
$\lambda g$ can be considered as an external potential.

The consistency of (\ref{h1}) in the limit
$G \rightarrow \infty$, is guaranteed by the closure of
the constraint algebra,

\begin{eqnarray}
&[& {\cal H}(x), {\cal H}(x^{'}) ] = 0, \nonumber \\
&[& {\cal H}(x), {\cal H}_i (x^{'}) ] =
( {\cal H} (x) + {\cal H} (x^{'})) ~ \delta_{,i} (x,x^{'}),
\label{ca} \\
&[& {\cal H}_i(x), {\cal H}_j (x^{'}) ] =
 {\cal H}_i (x) \delta_{,j} (x,x^{'}) +
 {\cal H}_j (x) \delta_{,i} (x,x^{'}). \nonumber
\end{eqnarray}

The strong coupling regime $G\gg1$ corresponds to a very high energy
region and when $G=\infty$ all the points of the spacetime become causally
disconnected, or equivalently, the Lorentz group is contracted to the
Carroll group \cite{levy}. Since this contraction is independent of the
value of $\lambda$, (\ref{ca}) corresponds to the contraction of both the
Lorentz and the anti-de Sitter group.

The limit $G\rightarrow \infty$ eliminates the spatial derivatives from
${\cal H}$. Thus, the dynamics of the gravitational field can only depend
on time, while the dependence on the space coordinates is only parametric.
This decoupling between space and time is particularly suited to describe
a homogeneous cosmology, where the time evolution of spatial sections
is a scaling transformation.

In cosmology one usually assumes isotropy and homogeneity of the spatial
sections. In the proper-time gauge, {\it i.e.} with the Lagrange
multipliers $N=g_{00}^{-1/2}$, $N_i = g_{0i}$ satisfying \cite{ct}
\begin{equation}
{\dot N} = 0, \,\,\,\,\,\,\,\,\,\,\,\,\, N_i=0, \label{pt}
\end{equation}
the space geometry takes the form
\begin{equation}
g_{ij} = f(t) \gamma_{ij}.
\end{equation}

Let us consider a topology of the spacetime of the form $T\times
\Sigma^{D-1}$ where $\Sigma^{D-1}$ is a closed compact surface.

In order to prove that (\ref{pt}) is really a good gauge, we should show
that (\ref{action}) is invariant under the gauge transformations generated
by ${\cal H}$ and ${\cal H}_i$,
{\it i.e.}
\begin{eqnarray}
\delta N &=& {\dot \epsilon} - {(N \epsilon^i )}_{,i}
+ {(N^i \epsilon)}_{,i}\nonumber\\
\delta N_i &=& {\dot \epsilon}_i.  \label{gt}
\end{eqnarray}
The operator in (\ref{gt}) is always
hyperbolic and can be inverted to solve for $\epsilon$. Replacing  (\ref{gt})
in the variation of (2) one finds
\begin{eqnarray}
\delta S &=& \delta \biggl(
\int d^Dx ( \pi^{ij} {\dot g_{ij}} - N {\cal H} - N^i {\cal H}_i )\biggr),
\nonumber\\
&=& \int d^{D-1}x ~ \biggl( \epsilon (G_{ijkl} \pi^{ij}
\pi^{kl} - \lambda g) + \epsilon^i {\cal H}_i
\biggr) \big|_{t_1}^{t_2} \nonumber
\\
&+& 2 \int dt \int_{\Sigma_{D-1}} d^{D-1} \sigma_l (N_k \delta \pi^{kl}).
\label{var}
\end{eqnarray}

In the first term, $\epsilon^i {\cal H}_i$ is cancelled because the
diffeomorphism constraints are satisfied,  while the other term is nonzero
unless

\begin{equation}
\epsilon (t_1, {\bf x}) = 0 = \epsilon (t_2, {\bf x}).
\label{c1}
\end{equation}
In addition to (\ref{c1}), one can impose the condition
\begin{equation}
\epsilon^i = 0, \,\,\, \forall ({\bf x}, t),
\label{c3}
\end{equation}
because there is no restriction on the tangential deformations.
These last requirements are consistent with the gauge conditions
(\ref{pt}) and establish that it is a good gauge condition  \cite{taylor}.

\section{Classical Cosmologies in the Strong Coupling Limit}

In this section we look for cosmological solutions to the Einstein
equations in the strong coupling limit for the topology $T\times
\Sigma^{D-1}$, where $\Sigma^{D-1}$ is a locally flat surface. Intuitively
one expects that the removal of spatial derivatives from the ${\cal H}$
constraint can be formally seen as a set of infinite massless free relativistic particles
moving in a $D$-dimensional spacetime.

Since in the strong coupling limit one neglects  the spatial derivatives,  a natural ansatz for
the metric is
\begin{equation}
g_{ij} = f(t) {\tilde g}_{ij},
\label{metric}
\end{equation}
where ${\tilde g}_{ij}$ is a function of ${\vec x}$ and, for $D=3$, could be
computed using the
uniformization theorem\cite{kra}. However, this is not necessary  because there are no
spatial derivatives in
$R^{(D-1)}$ and ${\tilde g}_{ij}$ only appears as a constant factor in the contraint
${\cal H}$ ({\it i.e.} only depending on ${\vec x}$).

The supermetric (\ref{smetric}) for this geometry becomes
\begin{equation}
G_{ijkl} =\frac{1}{2} f^2 \biggl({\tilde g} _{ik} {\tilde g}_{jl} +
{\tilde g}_{il} {\tilde g}_{jk} - \frac{2}{(D-2)} {\tilde g}_{ij} {\tilde g}_{kl}
\biggr),
\label{sm}
\end{equation}
and the Hamiltonian constraint is

\begin{eqnarray}
{\cal H} &=&  \frac{\tilde g ^{-1}(D-2)^3 (D-1)}{{(4 \pi N)}^2} f^{(1-D)}
{\biggl(\frac{{\dot
f}}{f}\biggr)}^2  \nonumber \\
               & & + \tilde g \lambda f^{(D-1)} = 0. \label{vinc}
\end{eqnarray}
Solving this equation one finds
\begin{equation}
f = \omega ~ t^{\frac{1}{1-D}}, \label{so}
\end{equation}
where
\begin{equation}
\omega = 4\pi N {\tilde g} \sqrt{\frac{-\lambda (D-1)}{(D-2)^3}}, \label{mendez}
\end{equation}

Looking at (\ref{mendez}) it is clear that the cosmological constant is
necessarily negative and from (\ref{vinc}) one see that the solution of the
equation of motion contains a time singularity as in standard 3+1 cosmology.

\section{ Quantum Mechanics in Superspace}

The aim of this section is to write down the propagation amplitude for three-
dimensional quantum gravity in the
strong
coupling limit which in the proper-time gauge reads
\begin{eqnarray}
G[g_{ij} (2), g_{ij} (1)] = \int {\cal D} N{\cal D} &g&_{ij} {\cal D}
\pi^{ij} \delta [ {\dot N}] \det (C)
\nonumber
\\
\times && e^{i \int d^3x ( \pi^{ij} {\dot
g_{ij}} - N {\cal H})}, \label{am1}
\end{eqnarray}
where $\det (C)$ is the Faddeev-Popov determinant  given by
\begin{equation}
\det (C) = \det \left(\begin{array}{ccl} \partial^2_{\tau} & \,\,\,\,\,\,\,N,_i
\partial_{\tau} +N\partial_{\tau}\partial_i \\
0 & \delta_{ij}\partial_{\tau} \end{array} \right).
\label{fp}
\end{equation}

However this expression remains formal as long as one does not
have an independent notion of time that could be used by an \lq\lq external" observer.
Technically speaking in
ordinary quantum mechanics the external time is identified with the direction in field
space which corresponds to a
negative contribution to the kinetic energy in the action \cite{isham}. This fact can be
easily visualized by considering  the free
massless relativistic particle described by the lagrangian
\begin{equation}
L = \frac{1}{2N} {\dot X}^\mu {\dot X}_\mu, \label{l1}
\end{equation}
where $N$ is the einbein. The temporal part of (\ref{l1}) is $-\frac{1}{2N} {\dot X}
_0^2$ and, futhermore, one
defines time as $X_0 = t$.

Our goal below will be find a similar structure for gravity.

In order to do that,  let us integrate $\pi^{ij}$ in (\ref{am1})
\begin{equation}
G[g_{ij} (2), g_{ij} (1)] = \int {\cal D} N {\cal D} g_{ij}
e^{ i\int d^3x (\frac{1}{2N} {\dot g}_{ij} G^{ijkl} {\dot g}_{kl}
- \frac{1}{2} \lambda N g )}, \label{am2}
\end{equation}
where the Faddeev-Popov determinant was absorved as an overall normalization.
The action in (\ref{am2}) has the structure ${\dot X}^A G_{AB} {\dot X}^B/2N$
if we identify
\begin{equation}
X_A \leftrightarrow g_{(ij)} \label{id}
\end{equation}
 where $N$  is the einbein and $\lambda N g$  plays the role of a scalar potential.
 Thus, (\ref{am2}) can be seen as
 describing a massless relativistic particle on a background metric $G$.

 However, as it is well known \cite{dewitt}, $G^{ijkl}= -\frac{1}{2}(g^{ik}g^{jl}+
 g^{il}g^{jk}
-2g^{ij}g^{kl})$ is the metric tensor over the  three-dimensional superspace with
signature $(-,+,+)$ and from here it is quite natural to interpret the volume of
the space  as time  in quantum gravity. This interpretation is very closed related to the functional difussion
equation of
p-brane theory \cite{eguchi} where for $p=0$ (particle) the time is the  proper-time , for  $p=1$ (string) it is the
area
of the world- sheet, for $p=2$ (2-brane)  it is the world-volume and so on.

Thus from this point of view the loop wave equation (for zero  cosmological constant) becomes
\begin{equation}
 - G^{ijkl} \frac{\delta^2}{\delta g^{ij} \delta g^{kl}} G[g_{ij} (2), g_{ij} (1)] = i \frac{\partial}{\partial V}
 G[g_{ij} (2), g_{ij} (1)], \label{wd}
 \end{equation}
 where in the RHS the Moser's theorem has been used\cite{moser}.

 It is interesting to observe that in the absence of a cosmological constant, three-dimensional quantum gravity in
 the strong coupling limit is exactly mapped to a tensionless 3-brane moving on a curved background.
Now, if we posit the correspondence\footnote{ For a related and interesting discussion see\cite{green}.}
\begin{equation}
M(g) \sim M(G), \label{prin}
\end{equation}
where $M(G)$ is the manifold of the target space and $\Sigma$ and $\Sigma^{\prime}$ are flat Riemann
surfaces in
$M(g)$  and $M(G)$. The propagation amplitude of geometries
becomes
\begin{equation}
G[g_{ij} (2), g_{ij} (1)] =  0, \label{cero}
\end{equation}
that is, there are no gravitons propagating.

This result subsists for any genus because the strong coupling limit does not permit the
formation of new geometrical
structures.

If the cosmological constant is different from zero,  the problem can be solved using the fact that
in two dimensions the topology
of the surface is classified by the genus, so that
\begin{equation}
G[g_{ij} (2), g_{ij} (1)] = \sum_{genus} G_{genus}[g_{ij} (2), g_{ij} (1)],
\label{uni}
\end{equation}
where $G_{genus}[g_{ij} (2), g_{ij} (1)]$ is the propagation amplitude for a given
genus. Each term of the sum may
be computed using the uniformization theorem \cite{kra}, but in the strong coupling
limit this is not necessary because the the world surface is  a set of infinite world-lines. Then after to use the
loop-wise expansion one has
\begin{eqnarray}
 G_{genus}[&g&_{ij} (2), g_{ij} (1)] =  \int_0^{\infty} {\cal D} N(x)~
 e^{-S_{cl}} ~  \int {\cal D} {h}_{ij} e^{-S({h})},
 \nonumber
 \\
 &=& \int_0^{\infty} {\cal D} N(x)~
 e^{-S_{cl}(g)}{ \det}^{- \frac{3}{2}} \biggl( \frac{\delta^2 S}{\delta g_{ij}(1)
 \delta g^{ij}(2)} \biggr) \label{final}
\end{eqnarray}
where $h_{ij}$ are quantum fluctuations,  the Pauli-de Witt-Van Vleck
determinant contains all the higher corrections in $\hbar$ and a Wick rotation is assumed.

Technically speaking, the Wick rotation is performed changing $N \rightarrow iN$ in (\ref{am2}) and one
obtain
\begin{equation}
iS_{cl} = -  \int d \tau [ \frac{1}{N} {(\frac{{\dot f}}{f})}^2  - \frac{1}{2}\lambda N \tilde g] \label{7}
\end{equation}
As a consequence, $e^{iS_{cl}}\rightarrow e^{-S_{cl}}$ and $S$ is positive definite if and only if the
cosmological
constant is negative. One should note also that $\tilde g > 0$ because we are considering oriented Riemann
surfaces
\cite{hat}. The conclusion $\Lambda<0$ again is obtained in agreement with (\ref{mendez}).

Using (\ref{so}) one finds
\begin{eqnarray}
 S_{cl} &=& \int_{\tau_1}^ {\tau_2} d \tau ( \frac{C_0}{\tau^2} + \frac{C_1}{\tau}), \label{action1}
 \\
 &=& - C_0 [\frac{1}{\tau_2} -   \frac{1}{\tau_1}] + C_1 ln (\frac{\tau_1}{\tau_2}), \label{action2}
 \end{eqnarray}
where $C_0$ and $C_1$ are constants that come from the spatial integration.
The final expression for the propagator is
 \begin{eqnarray}
 G_{genus} [g_{ij} (2), g_{ij} (1)] &=& \int_0^{\infty} {\cal D} N(x)~
\biggl(\frac{\tau_1}{\tau_2}\biggr)^{C_1} e^{C_0 ( \frac{1}{\tau_2} - \frac{1}{\tau_1})} \nonumber  \\
& &\times { \det}^{- \frac{3}{2}} \biggl( \frac{\delta^2 S}{\delta g_{ij}(1)  \delta g^{ij}(2)} \biggr) \label{gil}
\end{eqnarray}
one can note that  $G [g_{ij} (2), g_{ij} (1)]$  is different from zero for $\tau_1 \neq 0$ and zero for $\tau_1 =
0$
and therefore, quantum mechanically the big-bang singularity is supressed.

One could note that for $D>3$, (\ref{uni}) also vanishes implying the smoothing-out of the big-bang in higher
dimensions.

\section{Conclusions}

In conclusion, our results suggest no initial singularity for the universe. Probably at the beginning,  the
universe was regular but in unstable equilibrium and evolved to the present expansion by quantum fluctuations.

As a technical final note we would like to mention that when spatial corrections are included, the calculation of
(\ref{uni}) is more involved because the integration in $g_{ij}$ is highly non-trivial due to the no existence  of
a
theorem of classification of Riemann surfaces in higher dimensions. However in $D=3$ these corrections can be
taken in account -by using the uniformization theorem- and similar quantum effects for spatial
singularities (black-
holes) should be found. We are trying to understand this issue.

\acknowledgments

We would like to thank to J. Alfaro, E. Matute, F. Melo, R. Troncoso and J. Zanelli for useful discussions. This
work
was partially supported by grants 1980788, 3000005 from FONDECYT (Chile) and DICYT (USACH). One of
us (J. G.) thanks to Steve Carlip for discussions and hospitality at Davis and Fundaci\'on Andes for a fellowship.

\end{document}